\documentclass[conference]{IEEEtran}
\usepackage[T1]{fontenc}
\usepackage[latin9]{inputenc}
\usepackage{color}
\usepackage{float}
\usepackage{amsmath}
\usepackage{amssymb}
\usepackage{graphicx}
\usepackage{wasysym}

\makeatletter

\floatstyle{ruled}
\newfloat{algorithm}{tbp}{loa}
\providecommand{\algorithmname}{Algorithm}
\floatname{algorithm}{\protect\algorithmname}

\usepackage{algorithm}
\usepackage{algorithmic}

\usepackage{empheq}
\usepackage[compress]{cite}
\usepackage{eucal}
\usepackage{array}
\usepackage{cases}
\usepackage{mathrsfs}
\usepackage{mathrsfs}\usepackage[noblocks]{authblk}
\usepackage{amsfonts}
\allowdisplaybreaks[4]

\usepackage{babel}

\usepackage[noblocks]{authblk}

\makeatother

\usepackage{babel}
\begin{document}
\title{Error Propagation and Overhead Reduced Channel Estimation for RIS-Aided
Multi-User mmWave Systems}
\author{(Invited Paper)\\
$\text{Zhendong Peng}^{*}$, $\text{Cunhua Pan}^{\ddagger}$, $\text{Gui Zhou}^{\dagger}$,
$\text{Hong Ren}^{\ddagger}$\\
 $^{*}$University of Electronic Science and Technology of China,
China, $^{\ddagger}$ Southeast University, China,\\
 $^{\dagger}$Friedrich-Alexander-University Erlangen-N{\"u}rnberg , Germany\\
 Email: {Zhendong@std.uestc.edu.cn, \{cpan, hren\}@seu.edu.cn, gui.zhou@fau.de}}
\maketitle
\begin{abstract}
In this paper, we propose a novel two-stage based uplink channel estimation
strategy with reduced pilot overhead and error propagation for a reconfigurable
intelligent surface (RIS)-aided multi-user (MU) millimeter wave (mmWave)
system. Specifically, in Stage I, with the carefully designed RIS
phase shift matrix and introduced matching matrices, all users jointly
estimate the correlation factors between different paths of the common
RIS-base station (BS) channel, which achieves significant multi-user
diversity gain. Then, the inherent scaling ambiguity and angle ambiguity
of the mmWave cascaded channel are utilized to construct an ambiguous
common RIS-BS channel composed of the estimated correlation factors.
In Stage II, with the constructed ambiguous common RIS-BS channel,
each user uses reduced pilots to estimate their specific user-RIS
channel independently so as to obtain the entire cascaded channel.
The theoretical number of pilots required for the proposed method
is analyzed and the simulation results are presented to validate the
effectiveness of this strategy.
\end{abstract}


\section{Introduction}

Reconfigurable intelligent surface (RIS) technology is envisioned
to be a promising technique for enhancing the spectrum and energy
efficiency of 6G-and-beyond communications with relative low hardware
cost and energy consumption \cite{Pan_MAG,marco,Pan2019intelleget}.
To reap the benefits promised by RIS, accurate channel state information
(CSI) is required, which is challenging to achieve due to the lack
of complex signal process ability of the RIS. 

Recently, there have been many contributions on channel estimation
for RIS-aided communication systems. Most work focused mainly on single-user
system \cite{Overview_IEEE_LEE}, but it might be not appropriate
to apply the methods developed for this case to multi-user system
directly since the pilot overhead would be prohibitively large, which
is proportional to the number of users. On the other hand, for multi-user
system, the correlation relationship among different user's cascaded
channel has been exploited to enhance the estimation performance \cite{ris-omp-3}
and reduce the pilot overhead of channel estimation significantly
\cite{Liuliang-IRS,RIS_Error_Propagation,Zhou_ULA_TSP}. Specifically,
for unstructured channel models, other users use reduced pilots to
estimate their scaling coefficients with respect to the typical user
so as to obtain the corresponding cascaded channels \cite{Liuliang-IRS,RIS_Error_Propagation}.
Similarly, for structured channel models, other users' cascaded channel
can be estimated effectively with the re-parameterized common BS-RIS
channel, which is constructed based on the estimated typical user's
cascaded channel \cite{Zhou_ULA_TSP}. However, these three methods
mentioned above have a common issue that the existence of channel
estimation error from the typical user in the previous stage will
deteriorate the estimation accuracy of other users in the next stage,
which is known as error propagation. An optional method to suppress
this error propagation effect is the careful selection of the typical
user. The typical user can be chosen as the closest one to the RIS
for the less severe path loss. In addition, large number of pilots
are usually allocated to the typical user so as to ensure the more
accurate estimation performance of the typical user. However, these
operations introduce the extra complexity to the system and consume
excessive pilot overhead for the estimation of the typical user. Once
the estimated CSI of the typical user is inaccurate, the estimation
performance of the multi-user system will be severely degraded.

Motivated by the above, in this paper, we develop a two-stage based
uplink cascaded channel estimation strategy without choosing the typical
user for an RIS-aided millimeter wave (mmWave) multi-user system.
In Stage I, with the carefully designed RIS phase shift matrix and
the introduced matching matrices, all users jointly estimate the correlation
factors between different paths of the common RIS-BS channel, which
achieves multi-user diversity gain and suppresses the negative error
propagation impact. Then, we utilize the inherent scaling ambiguity
and angle ambiguity properties of the mmWave cascaded channel to construct
an ambiguous common RIS-BS channel composed of the obtained correlation
factors. In Stage II, based on the constructed common RIS-BS channel,
each user only uses reduced pilots to estimate their specific user-RIS
channel independently so as to obtain the full CSI of the cascaded
channel. Lastly, we analyze the theoretical minimum number of pilots
required for the strategy and present the corresponding simulation
results for the proposed method.

\textit{Notations}: For a matrix $\mathbf{A}$ of arbitrary size,
$\mathbf{A}^{*}$, $\mathbf{A}^{\mathrm{T}}$, $\mathbf{A}^{\mathrm{H}}$,
$\mathbf{A}^{\mathrm{\dagger}}$, and $\mathrm{vec}(\mathbf{A})$
stand for the conjugate, transpose, conjugate transpose, pseudo-inverse,
and vectorization of $\mathbf{A}$. The $m$-th row of $\mathbf{A}$
is denoted by $\mathbf{A}_{(m,:)}$. Additionally, the Khatri-Rao
product and Hadamard product between two matrices $\mathbf{A}$ and
$\mathbf{B}$ are denoted by $\mathbf{A}\diamond\mathbf{B}$ and $\mathbf{A}\odot\mathbf{B}$,
respectively. $\mathbf{I}$ and $\mathbf{0}$ denote an identity matrix
and an all-zero matrix with appropriate dimensions, respectively.
For a vector $\mathbf{a}$, $[\mathbf{a}]_{m:n}$ denotes the subvector
containing from the $m$-th element to the $n$-th element of $\mathbf{a}$,
respectively. The symbol $||\mathbf{a}||$ represents the Euclidean
norm of $\mathbf{a}$. $\mathrm{Diag}\{\mathbf{a}\}$ is a diagonal
matrix with the entries of $\mathbf{a}$ on its diagonal. The inner
product between two vectors $\mathbf{a}$ and $\mathbf{b}$ is denoted
by $\left\langle \mathbf{a},\mathbf{b}\right\rangle \triangleq\mathbf{a}^{\mathrm{H}}\mathbf{b}$.
$\mathrm{i}\triangleq\sqrt{-1}$ is the imaginary unit. $\mathbb{C}$
represents the set of complex numbers.

\section{System Model\label{sec:System-Model-and}}

We consider a narrow-band time-division duplex (TDD) mmWave system,
in which $K$ single-antenna users communicate with a BS equipped
with an $N$-antenna ULA. To improve the communication performance,
an RIS equipped with a passive reflecting ULA of dimension $M$ is
deployed. In addition, the direct channel between the BS and users
are assumed to be blocked.

\subsection{Transmission model}

Denote $\mathbf{H}\in\mathbb{C}^{N\times M}$ as the channel matrix
between the RIS and the BS, and $\mathbf{h}_{k}\in\mathbb{C}^{M\times1}$
as the channel matrix between user $k$ and the RIS, respectively.
The set of users is defined as $\mathcal{K}=\{1,\ldots,K\}$. Denote
$\mathbf{e}_{t}\in\mathbb{C}^{M\times1}$ as the phase shift vector
of the RIS in time slot $t$. During the uplink transmission, the
received signal from user $k$ at the BS in time slot $t$, is expressed
as 
\begin{align}
\mathbf{y}_{k}(t) & =\mathbf{H}\mathrm{Diag}\{\mathbf{e}_{t}\}\mathbf{h}_{k}\sqrt{P}s_{k}(t)+\mathbf{n}_{k}(t)\nonumber \\
 & =\mathbf{H}\mathrm{Diag}\{\mathbf{h}_{k}\}\mathbf{e}_{t}\sqrt{P}s_{k}(t)+\mathbf{n}_{k}(t)\nonumber \\
 & \triangleq\mathbf{G}_{k}\mathbf{e}_{t}\sqrt{P}s_{k}(t)+\mathbf{n}_{k}(t),\label{signal_model}
\end{align}
where $s_{k}(t)$ is the pilot symbol of user $k$ and $\mathbf{n}_{k}(t)\in\mathbb{C}^{N}\sim\mathcal{CN}(\mathbf{0},\delta^{2}\mathbf{I}_{N})$
is the corresponding additive white Gaussian noise (AWGN) with power
$\delta^{2}$. $P$ represents the transmit power. $\mathbf{G}_{k}=\mathbf{H}\mathrm{Diag}\{\mathbf{h}_{k}\}$
is the cascaded user-RIS-BS channel of user $k$, needed to be estimated
in this work. 

\subsection{Channel model\label{subsec:Channel-model}}

First, we consider a typical ULA with $Q$ elements, whose steering
vector $\mathbf{a}_{Q}(x)\in\mathbb{C}^{Q\times1}$ can be represented
by
\begin{equation}
\mathbf{a}_{Q}(x)=[1,e^{-\mathrm{i}2\pi x},\ldots,e^{-\mathrm{i}2\pi(Q-1)x}]^{\mathrm{T}}.\label{eq:ax}
\end{equation}
The variable $x$ is regarded as the spatial frequency and there exists
a one-to-one relationship between the spatial frequency $x$ and the
physical angle $\mathfrak{\varrho}$ as $x=\frac{d}{\lambda_{c}}\cos(\mathfrak{\varrho})$
when assuming $d\le\lambda_{c}/2$. Here, $\lambda_{c}$ is the carrier
wavelength and $d$ is the element spacing. In the remainder of the
paper we will refer to the angle information as spatial frequency
for simplicity. 

Due to the limited scattering characteristics in the mmWave environment,
we use the geometric channel model to rewrite the channel matrices
in (\ref{signal_model}), i.e., $\mathbf{H}$ and $\mathbf{h}_{k}$
as 
\begin{align}
\mathbf{H} & =\sum_{l=1}^{L}\alpha_{l}\mathbf{a}_{N}(\psi_{l})\mathbf{a}_{M}^{\mathrm{H}}(\omega_{l})=\mathbf{A}_{N}\boldsymbol{\Lambda}\mathbf{A}_{M}^{\mathrm{H}},\label{H_compact}\\
\mathbf{h_{\mathit{k}}} & =\sum_{j=1}^{J_{k}}\beta_{k,j}\mathbf{a}_{M}(\varphi_{k,j})=\mathbf{A}_{M,k}\boldsymbol{\beta}_{k},\forall k\in\mathcal{K},\label{hk_compact}
\end{align}
where $L$ and $J_{k}$ denote the number of propagation paths (scatterers)
between the BS and the RIS, and between the RIS and user $k$, respectively.
In addition, $\alpha_{l}$, $\psi_{l}$ and $\omega_{l}$ are the
complex path gain, AoA, and AoD of the $l$-th path in the common
RIS-BS channel, respectively. Similarly, $\beta_{k,j}$ and $\varphi_{k,j}$
represent the complex path gain and AoA of the $j$-th path in the
user $k$-RIS channel, respectively. Moreover, $\mathbf{A}_{N}=[\mathbf{a}_{N}(\psi_{1}),\ldots,\mathbf{a}_{N}(\psi_{L})]\in\mathbb{C}^{N\times L}$,
$\mathbf{A}_{M}=[\mathbf{a}_{M}(\omega_{1}),\ldots,\mathbf{a}_{M}(\omega_{L})]\in\mathbb{C}^{M\times L}$,
and $\boldsymbol{\Lambda}=\mathrm{Diag}\{\alpha_{1},\ldots,\alpha_{L}\}\in\mathbb{C}^{L\times L}$
are the AoA steering (array response) matrix, AoD steering matrix
and complex gain matrix of the common RIS-BS channel, respectively,
and $\mathbf{A}_{M,k}=[\mathbf{a}_{M}(\varphi_{k,1}),\ldots,\mathbf{a}_{M}(\varphi_{k,J_{k}})]\in\mathbb{C}^{M\times J_{k}}$
and $\boldsymbol{\beta}_{k}=[\beta_{k,1},\ldots,\beta_{k,J_{k}}]^{\mathrm{T}}\in\mathbb{C}^{J_{k}\times1}$
are the AoA steering matrix and complex gain vector of the specific
user $k$-RIS channel, respectively. 

Combining (\ref{H_compact}) and (\ref{hk_compact}), the cascaded
channel $\mathbf{G}_{k}$ in (\ref{signal_model}) can be rewritten
as 
\begin{equation}
\mathbf{G}_{k}=\mathbf{A}_{N}\boldsymbol{\Lambda}\mathbf{A}_{M}^{\mathrm{H}}\mathrm{Diag}\{\mathbf{A}_{M,k}\boldsymbol{\beta}_{k}\},\forall k\in\mathcal{K}.\label{cascaded_channel}
\end{equation}

\section{\textcolor{black}{Two-Stage Based Cascaded Channel Estimation for
a Multi-user System without Choosing a Typical User\label{sec:Estimation-of-the}}}

In this section, a two-stage based uplink cascaded channel estimation
strategy without choosing a typical user is proposed. Specifically,
in Stage I, an ambiguous common RIS-BS channel is constructed by all
users jointly so as to achieve multi-user diversity gain and suppress
the impact of error propagation. In Stage II, each user only needs
to estimate the specific user-RIS channel to obtain full CSI of the
cascaded channel. \textcolor{black}{Finally, the required pilot overhead
is analyzed.} 

\subsection{\textcolor{black}{Stage I: Estimation of} the Ambiguous Common RIS-BS
Channel}

In this subsection, we present the details on the estimation of the
ambiguous common RIS-BS channel, seen as a re-parameterized common
RIS-BS channel.

During Stage I, with the carefully designed RIS phase shift coefficients,
all users transmit the training pilots simultaneously so as to achieve
the multi-user diversity gain. Assume the pilot symbols satisfy $s_{k}(t)=1$
for $\forall k\in\mathcal{K}$, and the transmitted power of each
user $P$ is the same, which equals to $1$, so that the received
signal at the BS can be expressed as 
\begin{align}
\mathbf{y}(t) & =\mathbf{y}_{1}(t)+...+\mathbf{y}_{k}(t)+...+\mathbf{y}_{K}(t)+\mathbf{n}(t)\nonumber \\
 & =\mathbf{H}\mathrm{Diag}\{\mathbf{h}_{1}\}\mathbf{e}_{t}+...+\mathbf{H}\mathrm{Diag}\{\mathbf{h}_{K}\}\mathbf{e}_{t}+\mathbf{n}(t)\nonumber \\
 & =\mathbf{H}\mathrm{Diag}\{\mathbf{h}\}\mathbf{e}_{t}+\mathbf{n}(t),\label{received_signal}
\end{align}
where $\mathbf{h}\triangleq\sum_{k=1}^{K}\mathbf{h}_{k}$ is treated
as the equivalent user-RIS channel, $\mathbf{n}(t)\in\mathbb{C}^{N\times1}\sim\mathcal{CN}(0,\delta^{2}\mathbf{I})$
is the AWGN at the BS. Stacking $V$ time slots of (\ref{received_signal}),
the received matrix $\mathbf{Y}=\left[\mathbf{y}(1),\mathbf{y}(2),...,\mathbf{y}(V)\right]\in\mathbb{C}^{N\times V}$
is expressed as
\begin{align}
\mathbf{Y} & =\mathbf{H}\mathrm{Diag}\{\mathbf{h}\}[\mathbf{e}_{1},\mathbf{e}_{2},\ldots,\mathbf{e}_{V}]+[\mathbf{n}(1),\mathbf{n}(2),...,\mathbf{n}(V)]\nonumber \\
 & =\mathbf{A}_{N}\boldsymbol{\Lambda}\mathbf{A}_{M}^{\mathrm{H}}\mathrm{Diag}\{\mathbf{h}\}\mathbf{E}+\mathbf{N},\label{received matrix}
\end{align}
where $\mathbf{E}=\left[\mathbf{e}_{1},\mathbf{e}_{2}\ldots,\mathbf{e}_{V}\right]\in\mathbb{C}^{M\times V}$
is the RIS phase shift matrix during this stage and $\mathbf{N}_{k}=\left[\mathbf{n}(1),\ldots,\mathbf{n}(V)\right]\in\mathbb{C}^{N\times V}$.

\subsubsection{Common AoAs estimation\label{subsec:Common-AoAs-estimation}}

Estimating the AoA steering matrix of the common RIS-BS channel, i.e.,
$\mathbf{A}_{N}$, from (\ref{received matrix}) is a classical directional
of arrival (DOA) estimation problem in array processing, and can be
solved by many mature signal processing techniques \cite{Overview_IEEE_LEE}.
Due to the large scale antenna arrays employed at the BS with typically
$L\ll N$, the DFT-based method in \cite{Zhou_ULA_TSP} can be adopted
to obtain the estimate of the common AoA steering matrix. 

Denote the estimate of $\mathbf{A}_{N}$ as $\mathbf{\widehat{A}}_{N}=[\mathbf{a}_{N}(\widehat{\psi}_{1}),\ldots,\mathbf{a}_{N}(\widehat{\psi}_{L})]\in\mathbb{C}^{N\times L}$.
Due to the property that $\mathrm{rank}(\mathbf{\widehat{A}}_{N})=L$
\cite{CE_MIMO_RIS}, by substituting $\mathbf{A}_{N}=\mathbf{\widehat{A}}_{N}+\Delta\mathbf{A}_{N}$,
we can obtain the equivalent received matrix $\mathbf{\widehat{A}}_{N}^{\mathrm{\dagger}}\mathbf{Y}$
as
\begin{align}
\mathbf{\widehat{A}}_{N}^{\mathrm{\dagger}}\mathbf{Y}= & \boldsymbol{\Lambda}\mathbf{A}_{M}^{\mathrm{H}}\mathrm{Diag}\{\mathbf{h}\}\mathbf{E}+\mathbf{\widehat{A}}_{N}^{\mathrm{\dagger}}\bar{\mathbf{N}}\in\mathbb{C}^{L\times V},\label{equiv_received_MATRIX}
\end{align}
where $\Delta\mathbf{A}_{N}\triangleq\mathbf{A}_{N}-\mathbf{\widehat{A}}_{N}$
is treated as the estimation error between the common AoA and its
estimate, and $\bar{\mathbf{N}}\triangleq\mathbf{N}+\Delta\mathbf{A}_{N}\boldsymbol{\Lambda}\mathbf{A}_{M}^{\mathrm{H}}\mathrm{Diag}\{\mathbf{h}\}\mathbf{E}$
represents the equivalent noise matrix. 

\subsubsection{Correlation relationship between different paths}

With the estimated common AoA, i.e., $\mathbf{\widehat{A}}_{N}$,
a correlation relationship between different paths in the common RIS-BS
channel will be revealed, which helps us to construct the ambiguous
common RIS-BS channel. Specifically, calculating the $l$-th row and
the $r$-th row of $\mathbf{\widehat{A}}_{N}^{\mathrm{\dagger}}\mathbf{Y}$
in (\ref{equiv_received_MATRIX}), and then taking their conjugate
transpose, we have 
\begin{align}
[(\mathbf{\widehat{A}}_{N}^{\mathrm{\dagger}}\mathbf{Y})_{l,:}]^{\mathrm{H}} & =[(\boldsymbol{\Lambda}\mathbf{A}_{M}^{\mathrm{H}})_{l,:}\mathrm{Diag}\{\mathbf{h}\}\mathbf{E}]^{\mathrm{H}}+[(\mathbf{\widehat{A}}_{N}^{\mathrm{\dagger}}\bar{\mathbf{N}})_{l,:}]^{\mathrm{H}}\nonumber \\
 & =\mathbf{E}^{\mathrm{H}}\mathrm{Diag}\{\mathbf{h}^{*}\}\mathbf{a}_{M}(\omega_{l})\alpha_{l}^{*}+\mathbf{\tilde{n}}_{l},\label{l-th row}\\{}
[(\mathbf{\widehat{A}}_{N}^{\mathrm{\dagger}}\mathbf{Y})_{r,:}]^{\mathrm{H}} & =[(\boldsymbol{\Lambda}\mathbf{A}_{M}^{\mathrm{H}})_{r,:}\mathrm{Diag}\{\mathbf{h}\}\mathbf{E}]^{\mathrm{H}}+[(\mathbf{\widehat{A}}_{N}^{\mathrm{\dagger}}\bar{\mathbf{N}})_{r,:}]^{\mathrm{H}}\nonumber \\
 & =\mathbf{E}^{\mathrm{H}}\mathrm{Diag}\{\mathbf{h}^{*}\}\mathbf{a}_{M}(\omega_{r})\alpha_{r}^{*}+\mathbf{\tilde{n}}_{r},\label{r-th row}
\end{align}
where $\mathbf{\tilde{n}}_{l}\triangleq[(\mathbf{\widehat{A}}_{N}^{\mathrm{\dagger}}\bar{\mathbf{N}})_{l,:}]^{\mathrm{H}}$
and $\mathbf{\tilde{n}}_{r}\triangleq[(\mathbf{\widehat{A}}_{N}^{\mathrm{\dagger}}\bar{\mathbf{N}})_{r,:}]^{\mathrm{H}}$.
Apparently, the dominant terms of $[(\mathbf{\widehat{A}}_{N}^{\mathrm{\dagger}}\mathbf{Y})_{l,:}]^{\mathrm{H}}$
and $[(\mathbf{\widehat{A}}_{N}^{\mathrm{\dagger}}\mathbf{Y})_{r,:}]^{\mathrm{H}}$,
i.e., $\mathbf{E}^{\mathrm{H}}\mathrm{Diag}\{\mathbf{h}^{*}\}\mathbf{a}_{M}(\omega_{l})\alpha_{l}^{*}$
and $\mathbf{E}^{\mathrm{H}}\mathrm{Diag}\{\mathbf{h}^{*}\}\mathbf{a}_{M}(\omega_{r})\alpha_{r}^{*}$,
contain the whole information regarding the $l$-th path and the $r$-th
path, respectively, and there exists a relationship between these
two terms, reflecting the correlation between the corresponding two
paths. 

To illustrate this correlation relationship, define a matching matrix
$\mathbf{A_{\mathit{l}}}$ for the $l$-th path with respect to the
$r$-th path as
\begin{align}
\mathbf{A_{\mathit{l}}} & \triangleq\mathbf{A}(\varpi_{l})x_{l}=[\mathbf{U}_{V}\odot(\mathbf{a}_{V}(\varpi_{l})\mathbf{1_{\mathit{V}}^{\mathrm{T}}})]^{\mathrm{H}}\frac{1}{V}\mathbf{U}_{V}x_{l},\label{matching_matrix}
\end{align}
where $\mathbf{A}(\varpi_{l})\triangleq[\mathbf{U}_{V}\odot(\mathbf{a}_{V}(\varpi_{l})\mathbf{1_{\mathit{V}}^{\mathrm{T}}})]^{\mathrm{H}}\frac{1}{V}\mathbf{U}_{V}\in\mathbb{C}^{V\times V}$
is a complex nonlinear function of $\varpi_{l}$. $\mathbf{U}_{V}$
is a $V\times V$ DFT matrix with the $(n,m)$-th entry given by $[\mathbf{U}_{V}]_{n,m}=e^{-\mathrm{i}\frac{2\pi}{V}(n-1)(m-1)}$.
$\mathbf{a}_{V}(\varpi_{l})$ can be regarded as the array manifold
with dimension $V$ and $\mathbf{1}_{\mathit{V}}$ is an all-one vector
of size $V$. Here, the $r$-th path is treated as the reference path.\footnote{The reference index $r$ can be chosen based on the maximum received
power criterion, i.e., $r=\mathrm{arg}\max_{p\in[1,L]}||[(\mathbf{\widehat{A}}_{N}^{\mathrm{\dagger}}\mathbf{Y})_{p,:}]^{\mathrm{H}}||^{2}$.} $\varpi_{l}$ and $x_{l}$ are the rotation factor and scaling factor
for the $l$-th path with respect to the $r$-th path, respectively,
which are given by
\begin{equation}
\varpi_{l}=\omega_{r}-\omega_{l},~x_{l}=\frac{\alpha_{l}^{*}}{\alpha_{r}^{*}}.\label{rot_scale_fac}
\end{equation}
Clearly, $\varpi_{l}\in[-2\frac{d_{\mathrm{RIS}}}{\lambda_{c}},2\frac{d_{\mathrm{RIS}}}{\lambda_{c}}]$.
During this stage, the RIS phase shift matrix $\mathbf{E}$ in (\ref{received matrix})
needs to be designed carefully, which should satisfy the following
structure
\begin{equation}
\mathbf{E}=\left[\begin{array}{c}
\mathbf{U}_{V}^{\mathrm{H}}\ \brokenvert\ \mathbf{0_{\mathit{V\times(M-V)}}}\end{array}\right]^{\mathrm{H}}.\label{RIS_pattern}
\end{equation}
Where $\mathbf{U}_{V}\in\mathbb{C}^{V\times V}$ is a DFT matrix defined
in (\ref{matching_matrix}), satisfying the constant modulus constraint.
To show how the matching matrix $\mathbf{A_{\mathit{l}}}$ works intuitively,
the noise terms of (\ref{l-th row}) and (\ref{r-th row}) are momentarily
omitted. Then, we have 
\begin{align}
 & \mathbf{A_{\mathit{l}}}\mathbf{E}^{\mathrm{H}}\mathrm{Diag}\{\mathbf{h}^{*}\}\mathbf{a}_{M}(\omega_{r})\alpha_{r}^{*}\nonumber \\
= & [\mathbf{U}_{V}\odot(\mathbf{a}_{V}(\varpi_{l})\mathbf{1_{\mathit{V}}^{\mathrm{T}}})]^{\mathrm{H}}\frac{1}{V}\mathbf{U}_{V}x_{l}\mathbf{E}^{\mathrm{H}}\mathrm{Diag}\{\mathbf{h}^{*}\}\mathbf{a}_{M}(\omega_{r})\alpha_{r}^{*}\nonumber \\
= & [(\mathbf{1_{\mathit{V}}}\mathbf{a}_{V}^{\mathrm{H}}(\varpi_{l})\odot\mathbf{U}_{V}^{\mathrm{H}}][\mathbf{I_{\mathit{V}}}\brokenvert\mathbf{0_{\mathit{V\times(M-V)}}}]\mathrm{Diag}\{\mathbf{h}^{*}\}\mathbf{a}_{M}(\omega_{r})\alpha_{l}^{*}\nonumber \\
= & [\mathbf{I_{\mathit{V}}}\mathbf{U}_{V}^{\mathrm{H}}\mathrm{Diag}^{*}\{\mathbf{a}_{V}(\varpi_{l})\}\brokenvert\mathbf{0_{\mathit{V\times(M-V)}}}]\mathrm{Diag}\{\mathbf{h}^{*}\}\mathbf{a}_{M}(\omega_{r})\alpha_{l}^{*}\nonumber \\
= & \mathbf{E}^{\mathrm{H}}\mathrm{Diag}^{*}\{\mathbf{a}_{M}(\varpi_{l})\}\mathrm{Diag}\{\mathbf{h}^{*}\}\mathbf{a}_{M}(\omega_{r})\alpha_{l}^{*}\nonumber \\
= & \mathbf{E}^{\mathrm{H}}\mathrm{Diag}\{\mathbf{h}^{*}\}\mathrm{Diag}\{\mathbf{a}_{M}(-\varpi_{l})\}\mathbf{a}_{M}(\omega_{r})\alpha_{l}^{*}\nonumber \\
= & \mathbf{E}^{\mathrm{H}}\mathrm{Diag}\{\mathbf{h}^{*}\}\mathbf{a}_{M}(\omega_{l})\alpha_{l}^{*},\label{corr_relationship}
\end{align}
where the third equality is obtained using $(\mathbf{yx^{\mathrm{H}}\mathrm{)}\odot A=\mathrm{Diag}\{y\}A\mathrm{Diag}^{*}\{x\}}$.
From Eq. (\ref{corr_relationship}), the term $\mathbf{E}^{\mathrm{H}}\mathrm{Diag}\{\mathbf{h}^{*}\}\mathbf{a}_{M}(\omega_{l})\alpha_{l}^{*}$
can be expressed as the result of the term $\mathbf{E}^{\mathrm{H}}\mathrm{Diag}\{\mathbf{h}^{*}\}\mathbf{a}_{M}(\omega_{r})\alpha_{r}^{*}$
via the linear transformation $\mathbf{A_{\mathit{l}}}$. Thus the
correlation between the $l$-th path and the $r$-th path is depicted
by the two variables of matching matrix $\mathbf{A_{\mathit{l}}}$,
i.e., $\varpi_{l}$ and $x_{l}$. By analogy, we conclude that the
correlation relationship between any two paths in the common RIS-BS
channel can be described by their rotation factors and scaling factors.
Now we will show how to estimate these two kinds of factors, i.e.,
rotation factors and scaling factors, from the equivalent received
signal $\mathbf{\widehat{A}}_{N}^{\mathrm{\dagger}}\mathbf{Y}$.

\subsubsection{Estimation of the rotation factors and scaling factors\label{subsec:Estimation-of-the}}

We still take the $l$-th path and the $r$-th path, i.e., the reference
path, as an example to illustrate the method for estimating the rotation
factors and scaling factors. Similar to Eq. (\ref{corr_relationship}),
we process the received signal $[(\mathbf{\widehat{A}}_{N}^{\mathrm{\dagger}}\mathbf{Y})_{r,:}]^{\mathrm{H}}$
in (\ref{r-th row}) via the linear transformation $\mathbf{A_{\mathit{l}}}$
as
\begin{align}
\mathbf{A_{\mathit{l}}}[(\mathbf{\widehat{A}}_{N}^{\mathrm{\dagger}}\mathbf{Y})_{r,:}]^{\mathrm{H}} & =\mathbf{A_{\mathit{l}}}\mathbf{E}^{\mathrm{H}}\mathrm{Diag}\{\mathbf{h}^{*}\}\mathbf{a}_{M}(\omega_{r})\alpha_{r}^{*}+\mathbf{A_{\mathit{l}}}\mathbf{\tilde{n}}_{r}\nonumber \\
 & =\mathbf{E}^{\mathrm{H}}\mathrm{Diag}\{\mathbf{h}^{*}\}\mathbf{a}_{M}(\omega_{l})\alpha_{l}^{*}-\Delta\mathbf{\tilde{n}}_{r},\label{r-th row_LT}
\end{align}
where $\Delta\mathbf{\tilde{n}}_{r}\triangleq-\mathbf{A_{\mathit{l}}}\mathbf{\tilde{n}}_{r}$.
By replacing $\mathbf{E}^{\mathrm{H}}\mathrm{Diag}\{\mathbf{h}^{*}\}\mathbf{a}_{M}(\omega_{l})\alpha_{l}^{*}=\mathbf{A_{\mathit{l}}}[(\mathbf{\widehat{A}}_{N}^{\mathrm{\dagger}}\mathbf{Y})_{r,:}]^{\mathrm{H}}+\Delta\mathbf{\tilde{n}}_{r}$
and $\mathbf{A_{\mathit{l}}}=\mathbf{A}(\varpi_{l})x_{l}$, (\ref{l-th row})
is re-expressed as
\begin{align}
[(\mathbf{\widehat{A}}_{N}^{\mathrm{\dagger}}\mathbf{Y})_{l,:}]^{\mathrm{H}} & =\mathbf{E}^{\mathrm{H}}\mathrm{Diag}\{\mathbf{h}^{*}\}\mathbf{a}_{M}(\omega_{l})\alpha_{l}^{*}+\mathbf{\tilde{n}}_{l}\nonumber \\
 & =\mathbf{A}(\varpi_{l})x_{l}[(\mathbf{\widehat{A}}_{N}^{\mathrm{\dagger}}\mathbf{Y})_{r,:}]^{\mathrm{H}}+\mathbf{n}_{\mathrm{noise}}.\label{matching_Eq1}
\end{align}
Here, $\mathbf{n}_{\mathrm{noise}}\triangleq\Delta\mathbf{\tilde{n}}_{r}+\mathbf{\tilde{n}}_{l}$
represents the corresponding noise vector. Moreover, denote $[(\mathbf{\widehat{A}}_{N}^{\mathrm{\dagger}}\mathbf{Y})_{l,:}]^{\mathrm{H}}$
as $\mathbf{\tilde{y}}_{l}$ and $\mathbf{A}(\varpi_{l})[(\mathbf{\widehat{A}}_{N}^{\mathrm{\dagger}}\mathbf{Y})_{r,:}]^{\mathrm{H}}$
as $\mathbf{B}(\varpi_{l})\in\mathbb{C}^{V}$, (\ref{matching_Eq1})
can be written as
\begin{equation}
\mathbf{\tilde{y}}_{l}=\mathbf{B}(\varpi_{l})x_{l}+\mathbf{n}_{\mathrm{noise}}.\label{formula-y_l}
\end{equation}
So our aim is to estimate the rotation factor $\varpi_{l}$ and scaling
factor $x_{l}$ with the received $\mathbf{\tilde{y}}_{l}$, which
can be achieved by solving the problem shown below as
\begin{align}
(\varpi_{l},x_{l}) & =\mathrm{arg}\min_{\varpi\in[-2\frac{d_{\mathrm{RIS}}}{\lambda_{c}},2\frac{d_{\mathrm{RIS}}}{\lambda_{c}}],x}||\mathbf{\tilde{y}}_{l}-\mathbf{B}(\varpi)x||^{2},\label{SNL-LS}
\end{align}
where $\mathbf{B}(\varpi)$ is an extremely complex nonlinear function
with respect to $\varpi$. This problem is known as a separable nonlinear
least squares (SNL-LS) problem, in which the optimal $\varpi_{l}$
and $x_{l}$ can be found via the following steps.

First, when the $\varpi$ is fixed, the optimal $x$ is given by the
least square (LS) solution, i.e., $x_{\mathrm{opt}}=\mathbf{B}^{\dagger}(\varpi)\mathbf{\tilde{y}}_{l}$.
Then, by substituting $x_{\mathrm{opt}}$ into Problem (\ref{SNL-LS}),
the optimal $\varpi$ can be acquired by solving the new problem as
\begin{equation}
\varpi_{l}=\mathrm{arg}\min_{\varpi\in[-2\frac{d_{\mathrm{RIS}}}{\lambda_{c}},2\frac{d_{\mathrm{RIS}}}{\lambda_{c}}]}||\mathbf{\tilde{y}}_{l}-\mathbf{B}(\varpi)\mathbf{B}^{\dagger}(\varpi)\mathbf{\tilde{y}}_{l}||^{2},\label{SNL-LS-1}
\end{equation}
where the objective function of Problem (\ref{SNL-LS-1}), i.e., $||\mathbf{\tilde{y}}_{l}-\mathbf{B}(\varpi)\mathbf{B}^{\dagger}(\varpi)\mathbf{\tilde{y}}_{l}||^{2}$,
can be simplified as
\begin{align}
 & ||\mathbf{\tilde{y}}_{l}-\mathbf{B}(\varpi)\mathbf{B}^{\dagger}(\varpi)\mathbf{\tilde{y}}_{l}||^{2}\nonumber \\
= & \mathbf{\tilde{y}}_{l}^{\mathrm{H}}\mathbf{\tilde{y}}_{l}-\mathbf{\tilde{y}}_{l}^{\mathrm{H}}\mathbf{B}(\varpi)\mathbf{B}^{\dagger}(\varpi)\mathbf{\tilde{y}}_{l}\nonumber \\
= & \mathbf{\tilde{y}}_{l}^{\mathrm{H}}\mathbf{\tilde{y}}_{l}-\mathbf{\tilde{y}}_{l}^{\mathrm{H}}\mathbf{B}(\varpi)[\mathbf{B}^{\mathrm{H}}(\varpi)\mathbf{B}(\varpi)]^{-1}\mathbf{B}^{\mathrm{H}}(\varpi)\mathbf{\tilde{y}}_{l}\nonumber \\
= & \mathbf{\tilde{y}}_{l}^{\mathrm{H}}\mathbf{\tilde{y}}_{l}-\phi(\varpi)\left|\left\langle \mathbf{\tilde{y}}_{l},\mathbf{B}(\varpi)\right\rangle \right|^{2},\label{SNL-LS-Computation1}
\end{align}
where $\phi(\varpi)\triangleq[\mathbf{B}^{\mathrm{H}}(\varpi)\mathbf{B}(\varpi)]^{-1}$.
With Eq. (\ref{SNL-LS-Computation1}), Problem (\ref{SNL-LS-1}) is
equivalent to the optimization problem as
\begin{equation}
\varpi_{l}=\mathrm{arg}\max_{\varpi\in[-2\frac{d_{\mathrm{RIS}}}{\lambda_{c}},2\frac{d_{\mathrm{RIS}}}{\lambda_{c}}]}\phi(\varpi)\left|\left\langle \mathbf{\tilde{y}}_{l},\mathbf{B}(\varpi)\right\rangle \right|^{2}.\label{SNL-LS-2}
\end{equation}
For Problem (\ref{SNL-LS-2}), one-dimension search method can be
adopted to obtain the optimal $\varpi$, i.e., $\varpi_{l}$, whose
performance depends on the number of search grids. Once the rotation
factor $\varpi_{l}$ is obtained by solving Problem (\ref{SNL-LS-2}),
the scaling factor $x_{l}$ is given by 
\begin{equation}
x_{l}=\mathbf{B}^{\dagger}(\varpi_{l})\mathbf{\tilde{y}}_{l}.\label{scaling_opt}
\end{equation}

Finally, the rotation factors and the scaling factors for any paths
in the common RIS-BS channel with respect to the reference path, i.e.,
the $r$-th path, can be estimated. Specifically, for $\forall l\neq r$,
the rotation factors and the scaling factors between the $l$-th path
and the $r$-th path, i.e., $\varpi_{l}$ and $x_{l}$, are obtained
via the solution to Problem (\ref{SNL-LS-2}) and Eq. (\ref{scaling_opt}).
While for $l=r$, according to the definitions in (\ref{rot_scale_fac}),
the corresponding rotation factor and the scaling factor are given
by $\varpi_{r}=0$ and $x_{r}=1$. 

\subsubsection{Construction of the common RIS-BS Channel}

In previous subsections, we have shown that the rotation factors and
scaling factors, i.e., $\varpi_{l}$ and $x_{l}$ for $\forall l=\{1,2,...,L\}$,
can be estimated effectively by all users jointly with the designed
$\mathbf{E}$. Based on the obtained parameters, an ambiguous complex
gain matrix of the common RIS-BS channel, i.e., denoted by $\boldsymbol{\Lambda}_{\mathrm{s}}$,
and an ambiguous AoD steering matrix of the common RIS-BS channel,
i.e., denoted by $\mathbf{A}_{\mathrm{s}}$, are constructed as
\begin{align}
\boldsymbol{\Lambda}_{\mathrm{s}} & \triangleq\mathrm{Diag}\{x_{1}^{*},x_{2}^{*},\ldots,x_{L}^{*}\},\label{gain_matrix_common}\\
\mathbf{A}_{\mathrm{s}} & \triangleq[\mathbf{a}_{M}(-\varpi_{1}),\mathbf{a}_{M}(-\varpi_{2}),\ldots,\mathbf{a}_{M}(-\varpi_{L})].\label{AOD_matrix_common}
\end{align}
It is observed that the relationship between $\boldsymbol{\Lambda}_{\mathrm{s}}$
and $\boldsymbol{\Lambda}$, and the relationship between $\mathbf{A}_{\mathrm{s}}$
and $\mathbf{A}_{M}$ can be described as
\begin{align}
\boldsymbol{\Lambda}_{\mathrm{s}} & =\mathrm{Diag}\{\frac{\alpha_{1}}{\alpha_{r}},\frac{\alpha_{2}}{\alpha_{r}},\ldots,\frac{\alpha_{L}}{\alpha_{r}}\}=\frac{1}{\alpha_{r}}\boldsymbol{\Lambda},\label{gain_matrix_relationship}\\
\mathbf{A}_{\mathrm{s}} & =[\mathbf{a}_{M}(\omega_{1}-\omega_{r}),\mathbf{a}_{M}(\omega_{2}-\omega_{r}),\ldots,\mathbf{a}_{M}(\omega_{L}-\omega_{r})]\nonumber \\
 & =\mathrm{Diag}\{\mathbf{a}_{M}(-\omega_{r})\}\mathbf{A}_{M}.\label{AoD_matrix_relationship}
\end{align}

With the obtained $\mathbf{A}_{N}$, $\boldsymbol{\Lambda}_{\mathrm{s}}$
and $\mathbf{A}_{\mathrm{s}}$, the corresponding ambiguous common
RIS-BS channel, denoted by $\mathbf{H}_{\mathrm{s}}$, is naturally
constructed as 
\begin{align}
\mathbf{H}_{\mathrm{s}} & \triangleq\mathbf{A}_{N}\boldsymbol{\Lambda}_{\mathrm{s}}\mathbf{A}_{\mathrm{s}}^{\mathrm{H}}.\label{common RIS-BS}
\end{align}
Then, substituting $\boldsymbol{\Lambda}=\alpha_{r}\boldsymbol{\Lambda}_{\mathrm{s}}$
and $\mathbf{A}_{M}=\mathrm{Diag}\{\mathbf{a}_{M}(\omega_{r})\}\mathbf{A}_{\mathrm{s}}$
into $\mathbf{G}_{k}$ in (\ref{cascaded_channel}), we have 
\begin{align}
\mathbf{G}_{k}= & \mathbf{A}_{N}\boldsymbol{\Lambda}\mathbf{A}_{M}^{\mathrm{H}}\mathrm{Diag}\{\mathbf{A}_{M,k}\boldsymbol{\beta}_{k}\}\nonumber \\
= & \mathbf{A}_{N}(\alpha_{r}\boldsymbol{\Lambda}_{\mathrm{s}})\mathbf{A}_{\mathrm{s}}^{\mathrm{H}}\mathrm{Diag}\{\mathbf{a}_{M}(-\omega_{r})\}\mathrm{Diag}\{\mathbf{A}_{M,k}\boldsymbol{\beta}_{k}\}\nonumber \\
= & \mathbf{A}_{N}\boldsymbol{\Lambda}_{\mathrm{s}}\mathbf{A}_{\mathrm{s}}^{\mathrm{H}}\mathrm{Diag}\{(\mathrm{Diag}\{\mathbf{a}_{M}(-\omega_{r})\}\mathbf{A}_{M,k})(\alpha_{r}\boldsymbol{\beta}_{k})\}\nonumber \\
= & \mathbf{A}_{N}\boldsymbol{\Lambda}_{\mathrm{s}}\mathbf{A}_{\mathrm{s}}^{\mathrm{H}}\mathrm{Diag}\{\mathbf{A}_{\mathrm{s},k}\boldsymbol{\beta}_{\mathrm{s},k}\}\nonumber \\
= & \mathbf{H}_{\mathrm{s}}\mathrm{Diag}\{\mathbf{h}_{\mathrm{s},k}\},\forall k\in\mathcal{K},\label{GK_common}
\end{align}
where $\mathbf{A}_{\mathrm{s},k}\triangleq\mathrm{Diag}\{\mathbf{a}_{M}(-\omega_{r})\}\mathbf{A}_{M,k}=[\mathbf{a}_{M}(\varphi_{k,1}-\omega_{r}),\ldots,\mathbf{a}_{M}(\varphi_{k,J_{k}}-\omega_{r})]\in\mathbb{C}^{M\times J_{k}}$
and $\boldsymbol{\beta}_{\mathrm{s},k}\triangleq\alpha_{r}\boldsymbol{\beta}_{k}=[\alpha_{r}\beta_{k,1},\ldots,\alpha_{r}\beta_{k,J_{k}}]^{\mathrm{T}}\in\mathbb{C}^{J_{k}\times1}$
are the corresponding ambiguous AoA steering matrix and ambiguous
complex gain vector of the specific user-RIS channel for user $k$,
respectively. Accordingly, $\mathbf{h}_{\mathrm{s},k}\triangleq\mathbf{A}_{\mathrm{s},k}\boldsymbol{\beta}_{\mathrm{s},k}=\alpha_{r}\mathrm{Diag}\{\mathbf{a}_{M}(-\omega_{r})\}\mathbf{h}_{k}$
is the corresponding ambiguous specific user-RIS channel for user
$k$, that must still be found. In next subsection we will show how
to estimate $\mathbf{h}_{\mathrm{s},k}$ for $\forall k\in\mathcal{K}$
with the constructed $\mathbf{H}_{\mathrm{s}}$, leading to a significant
reduction in the pilot overhead.

\subsection{\textcolor{black}{Stage II: Estimation of}\textcolor{red}{{} }the Ambiguous
Specific User-RIS Channel\label{subsec:Stage-II:-Estimation}}

In Stage II, the users are required to transmit the pilot sequences
one by one for the estimation of the ambiguous specific user-RIS channel. 

Without loss of generality, we consider an arbitrary $k$ from $\mathcal{K}$
and show how to estimate user $k$'s ambiguous specific user-RIS channel,
i.e., $\mathbf{h}_{\mathrm{s},k}$. Assume $\tau_{k}$ pilots are
allocated for user $k$ in this stage. In addition, assume the pilot
symbols satisfy $s_{k}(t)=1$ and the transmitted power $P$ equals
to $1$ as before. Then, with Eq. (\ref{GK_common}), the received
signal from user $k$ at the BS in time slot $t$ can be expressed
as
\begin{align}
\mathbf{y}_{k}(t) & =\mathbf{G}_{k}\mathbf{e}_{t}+\mathbf{n}_{k}(t)\nonumber \\
 & =\mathbf{H}_{\mathrm{s}}\mathrm{Diag}\{\mathbf{h}_{\mathrm{s},k}\}\mathbf{e}_{t}+\mathbf{n}_{k}(t).\label{stageII_Tx_K}
\end{align}
For clear illustration, we still assume that the BS receives the pilot
sequence from time slot $1$ to time slot $\tau_{k}$, and thus the
received matrix $\mathbf{Y}_{k}=\left[\mathbf{y}_{k}(1),\ldots,\mathbf{y}_{k}(\tau_{k})\right]\in\mathbb{C}^{N\times\tau_{k}}$
during user $k$'s pilot transmission is expressed as
\begin{align}
\mathbf{Y}_{k} & =\mathbf{H}_{\mathrm{s}}\mathrm{Diag}\{\mathbf{h}_{\mathrm{s},k}\}\mathbf{E}_{k}+\mathbf{N}_{k}\nonumber \\
 & =\mathbf{A}_{N}\boldsymbol{\Lambda}_{\mathrm{s}}\mathbf{A}_{\mathrm{s}}^{\mathrm{H}}\mathrm{Diag}\{\mathbf{h}_{\mathrm{s},k}\}\mathbf{E}_{k}+\mathbf{N}_{k},\label{Tx_k_model}
\end{align}
where $\mathbf{E}_{k}=\left[\mathbf{e}_{1},\ldots,\mathbf{e}_{\tau_{k}}\right]\in\mathbb{C}^{M\times\tau_{k}}$
and $\mathbf{N}_{k}=\left[\mathbf{n}_{k}(1),\ldots,\mathbf{n}_{k}(\tau_{k})\right]\in\mathbb{C}^{N\times\tau_{k}}$. 

With the estimated common AoA steering matrix $\mathbf{\widehat{A}}_{N}$,
user $k$'s received matrix is processed similarly to what was done
for (\ref{equiv_received_MATRIX}) as
\begin{equation}
\mathbf{\widehat{A}}_{N}^{\mathrm{\dagger}}\mathbf{Y}_{k}=\boldsymbol{\Lambda}_{\mathrm{s}}\mathbf{A}_{\mathrm{s}}^{\mathrm{H}}\mathrm{Diag}\{\mathbf{h}_{\mathrm{s},k}\}\mathbf{E}_{k}+\mathbf{\widehat{A}}_{N}^{\mathrm{\dagger}}\bar{\mathbf{N}}_{k}\in\mathbb{C}^{L\times\tau_{k}},\label{equiv_received_MATRIX_K}
\end{equation}
where $\bar{\mathbf{N}}_{k}\triangleq\mathbf{N}_{k}+\Delta\mathbf{A}_{N}\boldsymbol{\Lambda}_{\mathrm{s}}\mathbf{A}_{\mathrm{s}}^{\mathrm{H}}\mathrm{Diag}\{\mathbf{h}_{\mathrm{s},k}\}$.
Then, vectorizing (\ref{equiv_received_MATRIX_K}) and defining $\mathbf{w}_{k}=\mathrm{vec}(\mathbf{\widehat{A}}_{N}^{\mathrm{\dagger}}\mathbf{Y}_{k})\in\mathbb{C}^{L\tau_{k}\times1}$,
we have 
\begin{align}
\mathbf{w}_{k} & =\mathrm{vec}(\boldsymbol{\Lambda}_{\mathrm{s}}\mathbf{A}_{\mathrm{s}}^{\mathrm{H}}\mathrm{Diag}\{\mathbf{h}_{\mathrm{s},k}\}\mathbf{E}_{k})+\mathrm{vec}(\mathbf{\widehat{A}}_{N}^{\mathrm{\dagger}}\bar{\mathbf{N}}_{k})\nonumber \\
 & =(\mathbf{E}_{k}^{\mathrm{T}}\diamond\boldsymbol{\Lambda}_{\mathrm{s}}\mathbf{A}_{\mathrm{s}}^{\mathrm{H}})\mathbf{h}_{\mathrm{s},k}+\bar{\mathbf{n}}_{k}\nonumber \\
 & =\mathbf{W}_{k}\mathbf{h}_{\mathrm{s},k}+\bar{\mathbf{n}}_{k},\label{Wk_vec}
\end{align}
where $\mathbf{W}_{k}\triangleq(\mathbf{E}_{k}^{\mathrm{T}}\diamond\boldsymbol{\Lambda}_{\mathrm{s}}\mathbf{A}_{\mathrm{s}}^{\mathrm{H}})\in\mathbb{C}^{L\tau_{k}\times M}$
and $\bar{\mathbf{n}}_{k}$ is the corresponding equivalent noise
vector given by $\mathrm{vec}(\mathbf{\widehat{A}}_{N}^{\mathrm{\dagger}}\bar{\mathbf{N}}_{k})\in\mathbb{C}^{L\tau_{k}\times1}$.
The second equality is obtained via $\mathrm{vec}(\mathbf{\mathbf{A}\mathrm{Diag}\{b\}C})=(\mathbf{\mathbf{C}^{\mathrm{T}}\diamond A})\mathbf{b}$. 

In particular, as illustrated in (\ref{GK_common}), $\mathbf{A}_{\mathrm{s},k}=[\mathbf{a}_{M}(\varphi_{k,1}-\omega_{r}),\ldots,\mathbf{a}_{M}(\varphi_{k,J_{k}}-\omega_{r})]$
where $\varphi_{k,j}-\omega_{r}$ for $\forall j\in\{1,...,J_{k}\}$
lies within $[-2\frac{d_{\mathrm{RIS}}}{\lambda_{c}},2\frac{d_{\mathrm{RIS}}}{\lambda_{c}}]$,
thus we can formulate (\ref{Wk_vec}) as a $J_{k}$-sparse signal
recovery problem:
\begin{align}
\mathbf{w}_{k} & =\mathbf{W}_{k}\mathbf{A}_{\mathrm{s},k}\boldsymbol{\beta}_{\mathrm{s},k}+\bar{\mathbf{n}}_{k}=\mathbf{W}_{k}\mathbf{A}\mathbf{d}_{k}+\bar{\mathbf{n}}_{k},\label{wk_sparse}
\end{align}
where $\mathbf{A}\mathbf{d}_{k}$ in the third equality is the virtual
angular domain (VAD) representation of $\mathbf{h}_{\mathrm{s},k}$.
$\mathbf{A}\in\mathbb{C}^{M\times D}$ is an overcomplete dictionary
matrix $(D\geq M)$, and the columns of $\mathbf{A}$ contain values
for $\mathbf{a}_{M}(\varphi_{k,j}-\omega_{r})$ on the angle grid.
$\mathbf{d}_{k}\in\mathbb{C}^{D\times1}$ is a sparse vector with
$J_{k}$ nonzero entries corresponding to the ambiguous gains $\{\alpha_{r}\beta_{k,j}\}_{j=1}^{J_{k}}$.
Accordingly, $\mathbf{w}_{k}$ is regarded as the equivalent measurement
vector for the estimation of $\mathbf{h}_{\mathrm{s},k}$. Hence,
the estimation problem of $\mathbf{h}_{\mathrm{s},k}$ corresponding
to (\ref{wk_sparse}) can be solved using CS-based techniques. To
obtain the best CS-based estimation performance, the RIS phase shift
matrix $\mathbf{E}_{k}$ should be designed to ensure that the columns
of the equivalent dictionary $\mathbf{W}_{k}\mathbf{A}$ are orthogonal.
A detailed design for $\mathbf{E}_{k}$ that achieves this goal can
be found in \cite{Zhou_ULA_TSP}. A simpler method is to choose $\mathbf{E}_{k}$
as the random Bernoulli matrix, i.e., randomly generate the elements
of $\mathbf{E}_{k}$ from $\{-1,+1\}$ with equal probability \cite{ris-omp-3}. 

To conclude, we obtained the estimate of $\mathbf{A}_{\mathrm{s},k}$
and $\boldsymbol{\beta}_{\mathrm{s},k}$ via the CS-based method,
and thus the estimate of ambiguous specific user-RIS channel, denoted
by $\hat{\mathbf{h}}_{\mathrm{s},k}$, can be obtained directly. Denote
the estimated common RIS-BS channel in (\ref{common RIS-BS}) as $\hat{\mathbf{H}}_{\mathrm{s}}$,
the estimate of the cascaded channel for user $k$ is given by $\widehat{\mathbf{G}}_{k}=\hat{\mathbf{H}}_{\mathrm{s}}\mathrm{Diag}\{\hat{\mathbf{h}}_{\mathrm{s},k}\}$.
Finally, the completed estimation of $\mathbf{G}_{k}$ for $1\leq k\leq K$
is summarized in Algorithm \ref{algorithm-1}.

\begin{algorithm}
\caption{Estimation of $\mathbf{G}_{k}$, $1\protect\leq k\protect\leq K$}

\label{algorithm-1}

\begin{algorithmic}[1]

\REQUIRE $\mathbf{Y}$ in (\ref{received matrix}), $\mathbf{Y}_{k}$
in (\ref{Tx_k_model}) for $1\leq k\leq K$.

\textbf{\textit{Stage I: }}\textit{Estimation of $\mathbf{H}_{\mathrm{s}}$.}

\STATE Obtain the estimate $\mathbf{\widehat{A}}_{N}$ via the DFT-based
method in \cite{Zhou_ULA_TSP};

\STATE Choose the reference path, denote its index as $r$;

\FOR{$1\leq l\leq L$, $l\neq r$}

\STATE Obtain $\hat{\varpi}_{l}$ and $\widehat{x}_{l}$ according
to (\ref{SNL-LS-2}) and (\ref{scaling_opt});

\ENDFOR

\STATE Construct the estimate $\widehat{\boldsymbol{\Lambda}}_{\mathrm{s}}$
and $\widehat{\mathbf{A}}_{\mathrm{s}}$ based on (\ref{gain_matrix_common})
and (\ref{AOD_matrix_common});

\STATE Obtain the estimate of the ambiguous common RIS-BS channel,
i.e., $\mathbf{\widehat{H}}_{\mathrm{s}}=\mathbf{\widehat{A}}_{N}\widehat{\boldsymbol{\Lambda}}_{\mathrm{s}}\widehat{\mathbf{A}}_{\mathrm{s}}^{\mathrm{H}}$;

\textbf{\textit{Stage II: }}\textit{Estimation of $\mathbf{h}_{\mathrm{s},k}$.}

\FOR{$1\leq k\leq K$}

\STATE Estimate the ambiguous specific user-RIS channel $\mathbf{h}_{\mathrm{s},k}$
from the sparse recovery problem associated with (\ref{wk_sparse});

\STATE Obtain the estimate of the cascaded channel, i.e., $\widehat{\mathbf{G}}_{k}=\widehat{\mathbf{H}}_{\mathrm{s}}\mathrm{Diag}\{\widehat{\mathbf{h}}_{\mathrm{s},k}\}$;

\ENDFOR

\ENSURE $\widehat{\mathbf{G}}_{k},1\leq k\leq K$.

\end{algorithmic}
\end{algorithm}

\subsection{Pilot Overhead Analysis}

Now we analyze the pilot overhead required for the proposed two-stage
based uplink channel emaciation strategy. For simplicity, $J_{1}=J_{2}=\cdots=J_{K}=J$
is assumed. 

In Stage I, the number of pilots is suggested to satisfy $V\geqslant L$
so as to ensure the vectors $\{\mathbf{a}_{V}(\varpi_{l})\}_{l=1}^{L}$
are linear independent. In Stage II, the number of pilots for user
$k$ directly affects the estimation performance for the sparse recovery
problem associated with (\ref{wk_sparse}), where the dimension of
the equivalent sensing matrix $\mathbf{W}_{k}\mathbf{A}$ is $L\tau_{k}\times D$
satisfying $D\geq M$, and the corresponding sparsity level is $J_{k}$.
To find a $l$-sparse complex signal with dimension $n$, the number
of measurements $m$ is on the order of $\mathcal{O}(l\log(n))$.
Therefore, user $k$ needs $\tau_{k}\geq\mathcal{O}(J_{k}\log(D)/L)\geq\mathcal{O}(J_{k}\log(M)/L)$
pilots. Consider $K$ users in total, the overall minimum pilot overhead
is $L+\mathcal{O}(KJ\log(M)/L)$. 

\section{Simulation Results\label{sec:Simulation-Results}}

In this section, we present several simulation results to validate
the effectiveness of the proposed method. The channel gains $\alpha_{l}$
and $\beta_{k,j}$ follow a complex Gaussian distribution with zero
mean and variance of $10^{-3}d_{\mathrm{BR}}^{-2.2}$ and $10^{-3}d_{\mathrm{RU}}^{-2.8}$,
respectively. Here, $d_{\mathrm{BR}}$, defined as the distance between
the BS and the RIS, is set to $10$ m, while, $d_{\mathrm{RU}}$,
defined as the distance between the RIS and the users, is assumed
to be $100$ m. The antenna spacing at the BS and the element spacing
at the RIS are set to $d_{\mathrm{BS}}=d_{\mathrm{RIS}}=\frac{\lambda_{c}}{2}$.
The number of paths between the BS and the RIS, and the number of
paths between the RIS and users are set to $L=5$ and $J_{1}=\cdots=J_{K}=4$,
respectively. The normalized mean square error (NMSE) is chosen as
the performance metric, which is defined by $\mathrm{NMSE}=\mathbb{E}\{(\sum_{k=1}^{K}||\widehat{\mathbf{G}}_{k}-\mathbf{G}_{k}||_{F}^{2})\mathbf{/}(\sum_{k=1}^{K}||\mathbf{G}_{k}||_{F}^{2})\}.$
We compare the proposed method with the following three channel estimation
methods: Direct-OMP Method \cite{ris-omp-1}, DS-OMP Method \cite{ris-omp-3},
and Typical User Required Method \cite{Zhou_ULA_TSP}. 

\begin{figure}
\begin{centering}
\includegraphics[width=0.75\columnwidth]{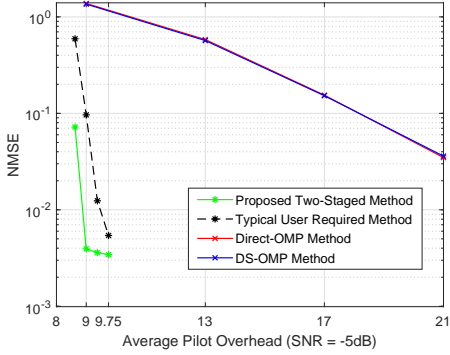}
\par\end{centering}
\caption{NMSEs vs. Average pilot overhead of each user $T$ when $N=100$,
$M=100$, $K=16$ and SNR = $-5$ dB.}

\label{pilot-ULA}
\end{figure}

Fig. \ref{pilot-ULA} illustrates the relationship between NMSE performance
and pilot overhead of the various methods, where the SNR is set to
$-5$ dB. Since different number of pilots are allocated in different
stages for the Proposed Two-Staged Method and the Typical User Required
Method, we consider the users' average pilot overhead, denoted as
$T$, and show NMSE as a function of $T$. It can be clearly seen
that an increase in the number of pilots improves the performance
of all algorithms. Under the same average pilot overhead, e.g., $T=9$,
the estimation performance of the Proposed Two-Staged Method markedly
outperforms than that of the other three methods due to the multi-user
diversity gain. On the other hand, in order to achieve the same estimation
performance, e.g., $\mathrm{NMSE}=10^{-2}$, the required average
pilot overhead of the Proposed Two-Staged Method and the Typical User
Required Method is much lower than the Direct-OMP Method and the DS-OMP
Method. That is because the former two methods exploit the correlation
relationship among different user's cascaded channel, which reduces
the pilot overhead.

\begin{figure}
\begin{centering}
\includegraphics[width=0.75\columnwidth]{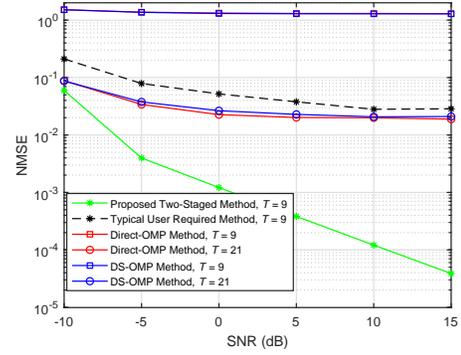}
\par\end{centering}
\caption{NMSEs vs. SNR for the ULA-type RIS case when $N=100$, $M=100$, $K=16$.}

\label{ULA-SNR}
\end{figure}

Fig. \ref{ULA-SNR} displays the NMSE performance of different methods
versus SNR. We observe the estimation performances of all the methods
are unacceptable under low SNR case, e.g., the NMSEs are larger than
$10^{-1}$ with SNR = $-10$ dB. Fortunately, their NMSEs are improved
with the growth of the SNR. In particular, the NMSE of the Proposed
Two-Staged Method decreases drastically with the SNR. By contrast,
due to the shortage of measurements, the estimation accuracy of the
three benchmark methods improves slightly and reaches saturation at
relatively high SNR. The gap between the proposed method and the three
benchmark methods becomes noticeably larger. 

\section{Conclusions\label{sec:Conclusions}}

In this paper, we proposed a novel two-stage based uplink channel
estimation strategy for an RIS-aided multi-user mmWave communication
system. In Stage I, all users jointly constructed the ambiguous common
RIS-BS channel so as to obtain multi-user diversity gain. In Stage
II, each user independently estimated their own ambiguous specific
user-RIS channel with reduced pilot overhead. Additionally, theoretical
overall minimum number of pilots required by the proposed strategy
was analyzed. Simulation results validated that the proposed method
outperforms other existing algorithms in terms of pilot overhead.

\bibliographystyle{IEEEtran}
\bibliography{bibfile}

\end{document}